# Efficiency of the principal component Liu-type estimator in logistic regression model

Jibo Wu[1] and Yasin Asar[2]

[1]*School of Mathematics and Finance, Chongqing University of Arts and Sciences, Chongqing, China*

[2]*Department of Mathematics-Computer Sciences, Necmettin Erbakan University, Konya, 42090, Turkey*

Jibo Wu:linfen52@126.com

Yasin Asar: yasar@konya.edu.tr, yasinasar@hotmail.com

# Efficiency of the principal component Liu-type logistic estimator in logistic regression model

In this paper we propose a principal component Liu-type logistic estimator by combining the principal component logistic regression estimator and Liu-type logistic estimator to overcome the multicollinearity problem. The superiority of the new estimator over some related estimators are studied under the asymptotic mean squared error matrix. A Monte Carlo simulation experiment is designed to compare the performances of the estimators using mean squared error criterion. Finally, a conclusion section is presented.

## 1. Introduction

Consider the following binary logistic regression model

$$\pi_i = \frac{\exp(x_i'\beta)}{1+\exp(x_i'\beta)}, i=1,...,n \qquad (1.1)$$

where $x_i' = (1\ x_{i1} \cdots x_{iq})$ denotes the $i$th row of $X$ which is an $n \times p\ (p = q+1)$ data matrix with $q$ known covariate vectors, $y_i$ shows the response variable which takes on the value either 0 or 1 with $y_i \sim Bernoulli(\pi_i)$, $y_i$'s are supposed to be independent of one another and $\beta' = (\beta_0\ \beta_1 \cdots \beta_q)$ stands for a $p \times 1$ vector of parameters.

Usually the maximum likelihood (ML) method is used to estimate $\beta$. The corresponding log-likelihood equation of model (1.1) is given by

$$L = \sum_{i=1}^{n} y_i \log(\pi_i) + (1-y_i)\log(1-\pi_i) \qquad (1.2)$$

where $\pi_i$ is the $i^{th}$ element of the vector $\pi$, $i = 1,2,...,n$.

ML estimator can be obtained by maximizing the log-likelihood equation given in (1.2). Since the equation (1.2) is non-linear in $\beta$, one should use an iterative algorithm called iteratively re-weighted least squares algorithm (IRLS) as follows (Saleh and Kibria, 2013) :

$$\hat{\beta}^{t+1} = \hat{\beta}^{t} + \left(X'V^{t}X\right)^{-1} X'V^{t}\left(y - \hat{\pi}^{t}\right) \quad (1.3)$$

where $\pi^{t}$ is the estimated values of $\pi$ using $\hat{\beta}_{t}$ and $V^{t} = \text{diag}\left(\hat{\pi}_{i}^{t}\left(1-\hat{\pi}_{i}^{t}\right)\right)$ such that $\hat{\pi}_{i}^{t}$ is the ith element of $\hat{\pi}_{t}$. After some algebra, Equation (1.3) can be written as follows:

$$\hat{\beta}_{ML} = \left(X'VX\right)^{-1} X'Vz \quad (1.2)$$

where $z' = (z_1 \cdots z_n)$ with $\eta_i = x_i'\beta$ and $z_i = \eta_i + (y_i - \pi_i)(\partial \eta_i / \partial \pi_i)$.

In linear regression analysis, multicollinearity has been regarded as a problem in the estimation. In dealing with this problem, many ways have been introduced to deal with this problem. One approach is to study the biased estimator such as ridge estimator (Hoerl and Kennard, 1970), Liu estimator (Liu, 1993), Liu-type estimator (Huang et al., 2009). Alternatively, many authors such as Xu and Yang (2011) and Li and Yang (2011), have studied the estimation of linear models with additional restrictions.

As in linear regression, estimation in logistic regression is also sensitive to multicollinearity. When there is multicollinearity, columns of the matrix $X'VX$ become close to be dependent. It implies that some of the eigenvalues of $X'VX$ become close to zero. Thus, mean squared error value of MLE is inflated so that one cannot obtain stable estimations. Thus many authors have studied how to reduce the multicollinearity, such as Lesaffre and Max (1993) discussed the multicollinearity in logistic regression, Schaefer et al. (1984) proposed the ridge logistic (RL) estimator, Aguilera et al. (2006) proposed the principal component logistic regression (PCLR) estimator, Masson et al. (2012), introduced the Liu logistic (LL) estimator, by combining the principal component logistic regression estimator and ridge logistic estimator to deal with multicollinearity. Moreover, Inan and Erdoğan (2013) proposed Liu-type logistic estimator (LTL) and Asar (2017) studied some properties of LTL.

In this study, by combining the principal component logistic regression estimator and the Liu-type logistic estimator, the principal component Liu-type logistic estimator is introduced as an alternative to the PCLR, ML and Liu-type logistic estimators to deal with the multicollinearity.

The rest of the paper is organized as follows. In Section 2, the new estimator and some properties of the new estimator are presented in Section 3. A Monte Carlo simulation is given in Section 4 and some concluding remarks are given in Section 5.

**2 The new estimator**

The logistic regression model is expressed by Aguilera et al. (2006) in matrix form in terms of the logit transformation as $L = X\beta = XTT'\beta = Z\alpha$ where $T = [t_1,...,t_p]$ shows an orthogonal matrix with $Z'VZ = T'X'VXT = \Lambda$ and $\Lambda = diag(\lambda_1,...,\lambda_p)$, $\lambda_1 \geq ... \geq \lambda_p$ is the ordered eigenvalues of $X'VX$. Then T and $\Lambda$ may be written as $T = (T_r\ T_{p-r})$ and $\begin{bmatrix} \Lambda_r & O \\ O & \Lambda_{p-r} \end{bmatrix}$ where $Z'_r VZ_r = T'_r X'VXT_r = \Lambda_r$ and $Z'_{p-r}VZ_{p-r} = T'_{p-r}X'VXT_{p-r} = \Lambda_{p-r}$. The Z matrix and the $\alpha$ vector can be partitioned as $Z = (Z_r\ Z_{p-r})$ and $\alpha = (\alpha'_r\ \alpha'_{p-r})'$. The handling of multicollinearity by means of PCR corresponds to the transition from the model $L = X\beta = XT_rT'_r\beta + XT_{p-r}T'_{p-r}\beta = Z_r\alpha_r + Z_{p-r}\alpha_{p-r}$ to the reduced model $L = Z_r\alpha_r$. The by equation (1) and PCR method we get the PCR estimator.

Inan and Erdoğan (2013) proposed Liu-type logistic estimator (LTL)

$$\hat{\beta}(k,d) = (X'VX + kI)^{-1}(X'Vz - d\hat{\beta}_{ML}) \tag{2.1}$$

where $-\infty < d < \infty$ and $k > 0$ are biasing parameters.

The principal component logistic regression estimator (Aguilera et al., 2006) is defined as

$$\hat{\beta}_r = T_r \left(T_r' X' V X T_r\right)^{-1} T_r' X' V z \tag{2.2}$$

We can write (2.2) as follows:

$$\hat{\beta}_r = T_r \left(T_r' X' V X T_r\right)^{-1} T_r' X' V z = T_r T_r' \hat{\beta}_{ML} \tag{2.3}$$

Then we can introduce a new estimator by replacing $\hat{\beta}^*(k,d)$ with $\hat{\beta}_{ML}$ in (2.3), and we get

$$\hat{\beta}_r(k,d) = T_r T_r' \hat{\beta}(k,d) = T_r \left(T_r' X' V X T_r + kI_r\right)^{-1} \left(T_r' X' V X T_r - dI_r\right) \left(T_r' X' V X T_r\right)^{-1} T_r' X' V z \tag{2.4}$$

where $-\infty < d < \infty$ and $k > 0$ are biasing parameters. We call this estimator as principal component Liu-type logistic regression (PCLTL) estimator.

**Remark:**

(1) It is obvious that $\hat{\beta}_r(k,d) = T_r \left(T_r' X' V X T_r + kI_r\right)^{-1} \left(T_r' X' V X T_r - dI_r\right) T_r' \hat{\beta}_r$, thus we can see the PCLTL estimator is a linear combination of the PCLR estimator.

(2) It is easy to obtain

(a) $\hat{\beta}_r(0,0) = \hat{\beta}_r = T_r \left(T_r' X' V X T_r\right)^{-1} T_r' X' V z$, PCLR estimator

(b) $\hat{\beta}_p(0,0) = \hat{\beta}_{ML} = \left(X' V X\right)^{-1} X' V z$, ML estimator

(c) $\hat{\beta}_p(k,d) = \hat{\beta}(k,d) = (X'VX + kI)^{-1}(X'Vz - d\hat{\beta}_{ML})$, LTL estimator.

Thus, the new estimator in (2.4) includes the PCLR, ML and LTL estimators as its special cases.

The next section we will study the properties of the new estimator.

## 3. The properties of the new estimator

For the sake of convenience, we show some lemmas which are needed in the following discussions.

**Lemma 3.1.** (Farebrother, 1976, Rao and Tountenburg, 1995) Suppose that $M$ be a positive definite matrix, namely $M > 0$, $\alpha$ be some vector, then $M - \alpha\alpha' \geq 0$ if and only if $\alpha' M^{-1}\alpha \leq 1$.

**Lemma 3.2.** (Baksalary and Trenkler, 1991) Let $C_{n \times p}$ be the set of complex matrices and $H_{n \times n}$ be the Hermitian matrices. Further, given $L \in C_{n \times p}$, $L^*$, $R(L)$ and $\kappa(L)$ denote the conjugate transpose, the range and the set of all generalized inverses, respectively of $L$. Let $A \in H_{n \times n}$, $a_1 \in C_{n \times 1}$ and $a_2 \in C_{n \times 1}$ be linearly independent, $f_{ij} = a_i' A^- a_j, i, j = 1, 2$ and $A \in \kappa(L)$, $a_1 \notin R(A)$. Let

$$s = \left[ a_1'(I - AA^-)'(I - AA^-) a_2 \right] \Big/ \left[ a_1'(I - AA^-)'(I - AA^-) a_1 \right]$$

Then $A + a_1 a_1' - a_2 a_2' \geq 0$ if and only if one of the following sets of conditions holds:

(a) $A \geq 0$, $a_i \in R(A)$, $i = 1, 2$, $(f_{11} + 1)(f_{22} - 1) \leq |f_{12}|^2$

(b) $A \geq 0$, $a_1 \notin R(A)$, $a_2 \in R(A : a_1)$, $(a_2 - sa_1)' A^- (a_2 - sa_1) \leq 1 - |s|^2$

(c) $A = U\Delta U' - \lambda vv'$, $a_i \in R(A)$, $i = 1, 2$, $v'a_1 \neq 0$, $f_{11} + 1 \leq 0$, $f_{22} - 1 \leq 0$, $(f_{11} + 1)(f_{22} - 1) \geq |f_{12}|^2$,

where $(U:v)$ shows a sub-unitary matrix, $\lambda$ shows a positive scalar. $\Delta$ shows a positive definite diagonal matrix. Further, the condition (a), (b) and (c) denote all independent of the choice of $A^-$, $A^-$ stands for the generalized inverse of A.

To compare the estimators, we use the mean squared error matrix (MSEM) criterion which is defined for an estimator $\check{\beta}$ as follows:

$$MSEM(\check{\beta}) = Cov(\check{\beta}) + Bias(\check{\beta})Bias(\check{\beta})'$$

where $Cov(\check{\beta})$ is the covariance matrix of $\check{\beta}$, and $Bias(\check{\beta})$ is the bias vector of $\check{\beta}$. Moreover, scalar mean squared error (SMSEM) of an estimator $\check{\beta}$ is also given as

$$SMSE(\check{\beta}) = tr\{MSEM(\check{\beta})\}.$$

### *3.1 Comparison of the new estimator (PCLTL) to the ML estimator*

For (2.4), we can compute the asymptotic variance of the new estimator as follows:

$$Cov\left(\hat{\beta}_r(k,d)\right) = T_r S_r(k)^{-1} \Lambda_r^{-1} S_r(d) \Lambda_r S_r(d) \Lambda_r^{-1} S_r(k)^{-1} T_r' \qquad (3.1)$$

where $S_r(k) = \Lambda_r + kI_r$, $S_r(d) = \Lambda_r - dI_r$.

Using (2.4), we get:

$$E\left(\hat{\beta}_r(k,d)\right) = T_r S_r(k)^{-1} \Lambda_r^{-1} S_r(d) \Lambda_r T_r' \beta \qquad (3.2)$$

By

$$T_r S_r(k)^{-1} \Lambda_r T_r' - I_p = -\left(T_{p-r} T_{p-r}' + k T_r S_r(k)^{-1} T_r'\right) \qquad (3.3)$$

Then we get the asymptotic bias of the new estimator as follows:

$$Bias\left(\hat{\beta}_r(k,d)\right) = \left(-T_{p-r} T_{p-r}' - (d+k) T_r S_r(k)^{-1} T_r'\right)\beta$$

We can get the asymptotic mean squared error matrix of the new estimator as follows

$$MSEM\left(\hat{\beta}_r(k,d)\right) = T_r S_r(k)^{-1} \Lambda_r^{-1} S_r(d) \Lambda_r S_r(d) \Lambda_r^{-1} S_r(k)^{-1} T_r'$$

$$+ \left(-T_{p-r} T_{p-r}' - (d+k) T_r S_r(k)^{-1} T_r'\right) \beta$$

$$\times \beta' \left(-T_{p-r} T_{p-r}' - (d+k) T_r S_r(k)^{-1} T_r'\right) \quad (3.4)$$

**Theorem 3.1.** Assume that $d < k$ and $d + k > 0$, then the new estimator is superior to the ML estimator under the asymptotic mean squared error matrix criterion if and only if

$$\beta' T_r (k+d)^2 \left[2(k+d)I_r + (k^2 - d^2)\Lambda_r^{-1}\right]^{-1} T_r' \beta + \beta' T_{p-r} \Lambda_{p-r} T_{p-r}' \beta \leq 1.$$

**Proof:** The asymptotic mean squared error matrix of ML estimator

$$MSEM\left(\hat{\beta}\right) = (X'VX)^{-1}. \quad (3.5)$$

By $\Lambda = \begin{pmatrix} \Lambda_r & O \\ O & \Lambda_{p-r} \end{pmatrix}$ and $T = (T_r, T_{p-r})$, we may obtain

$$(X'VX)^{-1} = T\Lambda^{-1}T' = T_r \Lambda_r^{-1} T_r' + T_{p-r} \Lambda_{p-r}^{-1} T_{p-r}'. \quad (3.6)$$

Let us consider the following difference

$$MSEM\left(\hat{\beta}\right) - MSEM\left(\hat{\beta}_r(k,d)\right)$$

$$= T_r S_r(k)^{-1} \left[2(k+d)I_r + (k^2 - d^2)\Lambda_r^{-1}\right] S_r(k)^{-1} T_r'$$

$$+ T_{p-r} \left[\Lambda_{p-r} - T_{p-r}' \beta \beta' T_{p-r}\right] T_{p-r}' - (k+d)^2 T_r S_r(k)^{-1}$$

$$\times T_r' \beta \beta' S_r(k)^{-1} T_r' + (k+d) T_r S_r(k)^{-1} T_r' \beta \beta' T_{p-r} T_{p-r}'$$

$$+ (k+d) T_{p-r} T_{p-r}' \beta \beta' T_r S_r(k)^{-1} T_r'. \quad (3.7)$$

Let

$$S^* = \begin{pmatrix} \dfrac{S_r(k)}{k+d} & 0 \\ 0 & \Lambda_{p-r} \end{pmatrix} \quad (3.8)$$

$$(\Lambda^*)^{-1} = \begin{pmatrix} \dfrac{2(k+d)I_r + (k^2-d^2)\Lambda_r^{-1}}{(k+d)^2} & 0 \\ 0 & \Lambda_{p-r} \end{pmatrix} \quad (3.9)$$

Now we can write (3.7) as

$$MSEM(\hat{\beta}) - MSEM(\hat{\beta}_r(k,d)) = T(S^*)^{-1}\left[(\Lambda^*)^{-1} - T'\beta\beta'T\right](S^*)^{-1}T' \quad (3.10)$$

Thus $MSEM(\hat{\beta}) - MSEM(\hat{\beta}_r(k,d))$ is a nonnegative definite matrix if and only if $(\Lambda^*)^{-1} - T'\beta\beta'T$ is a nonnegative definite matrix. Using Lemma 3.1, $(\Lambda^*)^{-1} - T'\beta\beta'T$ is a nonnegative definite matrix if and only if $\beta'T\Lambda^*T'\beta \leq 1$. Invoking the notation of $\Lambda^*$ in (3.9), we can prove Theorem 3.1.

### 3.2 Comparison of the new estimator (PCLTL) to the PCLR estimator

**Theorem 3.2.** Suppose that $d < k$ and $d + k > 0$, then the new estimator is better than the PCLR estimator under the asymptotic mean squared error matrix criterion if and only if $T_r'\beta = 0$.

**Proof:** Suppose that $k = d$ in Equation (3.4), then we get

$$MSEM(\hat{\beta}_r) = T_r\Lambda_r^{-1}T_r' + (T_rT_r' - I_p)\beta\beta'(T_rT_r' - I_p) \quad (3.11)$$

Now let us consider

$$MSEM(\hat{\beta}_r) - MSEM(\hat{\beta}_r(k,d))$$

$$= T_r S_r(k)^{-1}\left[2(k+d)I_r + (k^2 - d^2)\Lambda_r^{-1}\right]S_r(k)^{-1}T_r'$$

$$+ (T_rT_r' - I_p)\beta\beta'(T_rT_r' - I_p)$$

$$+ \left(-T_{p-r}T_{p-r}' - (d+k)T_rS_r(k)^{-1}T_r'\right)\beta$$

$$\times \beta'\left(-T_{p-r}T_{p-r}' - (d+k)T_rS_r(k)^{-1}T_r'\right) \quad (3.12)$$

To apply Lemma 3.2, let $A = T_r B T_r'$, where

$$B = S_r(k)^{-1}\left[2(k+d)I_r + (k^2-d^2)\Lambda_r^{-1}\right]S_r(k)^{-1} \quad (3.13)$$

And $a_1 = (T_r T_r' - I_p)\beta$, $a_2 = \left(-T_{p-r}T_{p-r}' - (d+k)T_r S_r(k)^{-1}T_r'\right)\beta$.

When $d < k$ and $d + k > 0$, B is a positive definite matrix. Then we get the Moore-Penrose inverses of $A$ is $A^+ = T_r B^{-1} T_r'$, and $AA^+ = T_r T_r'$. Thus $a_1 \in R(A)$ if and only if $a_1 = 0$. Since $a_1 \neq 0$, we cannot use part (a) and (c) of Lemma 3.2, we can only apply part (b) of Lemma 3.2. Using the definition of $s$, we may obtain that $s = 1$. On the other hand, $a_2 - a_1 = A\eta$, where

$$\eta = (d+k)T_r S_r(k)\left[2(k+d)I_r + (k^2-d^2)\Lambda_r^{-1}\right]^{-1}T_r'\beta \quad (3.14)$$

Thus, we can easily obtain $a_2 \in R(A : a_1)$. Then Using Lemma 3.2, we can get that the new estimator is superior to the PCLR estimator under the asymptotic mean squared error matrix criterion if and only if $(a_2 - a_1) A^-(a_2 - a_1) \leq 0$ or $\eta' A \eta \leq 0$. In fact, $(a_2 - a_1) A^-(a_2 - a_1) \geq 0$, so the new estimator is better than the PCLR estimator under the asymptotic mean squared error matrix criterion if and only if $\eta' A \eta = 0$, that is

$$\beta' T_r \left[2(k+d)I_r + (k^2-d^2)\Lambda_r^{-1}\right]^{-1} T_r'\beta = 0 \quad (3.15)$$

And $\beta' T_r \left[2(k+d)I_r + (k^2-d^2)\Lambda_r^{-1}\right]^{-1} T_r'\beta = 0$ if and only if $T_r'\beta = 0$.

### 3.3 Comparison of the new estimator (PCLTL) to the Liu-type logistic estimator

**Theorem 3.3.** The new estimator is superior to the Liu-type logistic estimator under the asymptotic mean squared error matrix criterion if and only if $T_{p-r}'\beta = 0$.

Proof: Put $r = p$ into (3.4), we get

$$MSEM\left(\hat{\beta}(k,d)\right) = TS(k)^{-1}S(d)\Lambda^{-1}S(d)S(k)^{-1}T'$$

$$+(k+d)^2 TS(k)^{-1} T' \beta\beta' TS(k)^{-1} T' \qquad (3.16)$$

Where $S(k) = \Lambda + kI_p$ and $S(d) = \Lambda - dI_p$.

Now we study the following difference

$$MSEM\left(\hat{\beta}(k,d)\right) - MSEM\left(\hat{\beta}_r(k,d)\right)$$

$$= TS(k)^{-1} S(d) \Lambda^{-1} S(d) S(k)^{-1} T'$$

$$- T_r S_r(k)^{-1} \Lambda_r^{-1} S_r(d) \Lambda_r S_r(d) \Lambda_r^{-1} S_r(k)^{-1} T_r'$$

$$+ (k+d)^2 TS(k)^{-1} T' \beta\beta' TS(k)$$

$$- \left(-T_{p-r} T'_{p-r} - (d+k) T_r S_r(k)^{-1} T_r'\right) \beta$$

$$\times \beta' \left(-T_{p-r} T'_{p-r} - (d+k) T_r S_r(k)^{-1} T_r'\right) \qquad (3.17)$$

Suppose that $C = T_{p-r} D T'_{p-r}$, where

$$D = S_{p-r}(k)^{-1} S_{p-r}(d) \Lambda_{p-r}^{-1} S_{p-r}(d) S_{p-r}(k)^{-1}$$

and $a_3 = (d+k)TS(k)^{-1} T' \beta$, $a_2 = \left(-T_{p-r} T'_{p-r} - (d+k) T_r S_r(k)^{-1} T_r'\right)\beta$. We can apply part (b) of Lemma 3.2. The Moore-Penrose inverse of $C$ is $C^+ = T_{p-r} D^{-1} T'_{p-r}$, and $CC^+ = T_{p-r} T'_{p-r}$. So $a_3 \notin R(C)$, $a_2 \in R(C:a_3)$, $s=1$ and $a_2 - a_3 = C\eta_1$, where

$$\eta_1 = -T_{p-r} S_{p-r}(k)^{-1} S_{p-r}(d) \Lambda_{p-r}^{-1} T'_{p-r} \beta$$

Then by Lemma 3.2, we obtain the new estimator is superior to the Liu-type logistic estimator under the asymptotic mean squared error matrix criterion if and only if $(a_2 - a_3) C^-(a_2 - a_3) \leq 0$ or $\eta_1' C \eta_1 \leq 0$. In fact, $(a_2 - a_3) C^-(a_2 - a_3) \geq 0$, so the new estimator is better than the Liu-type logistic estimator under the asymptotic mean squared error matrix criterion if and only if $\eta_1' C \eta_1 = 0$, that is $\beta' T_{p-r} \Lambda_{p-r} T'_{p-r} \beta = 0$.

## 4. Monte Carlo Simulation Study

In this simulation study, we study the logistic regression model. In this section, we present the details and the results of the Monte Carlo simulation which is conducted to evaluate the performances of the estimators MLE, PCLR, and LTL estimators and PCLTL. There are several papers studying the performance of different estimators in the binary logistic regression. Therefore, we follow the idea of Lee and Silvapulle (1988), Månsson, Kibria and Shukur (2012), Asar (2017) and Asar and Genç (2016) generating explanatory variables as follows

$$x_{ij} = \left(1-\rho^2\right)^{1/2} z_{ij} + \rho z_{iq+1} \qquad (4.1)$$

where $i=1,2,...,n$, $j=1,2,...,q$ and $z_{ij}$'s are random numbers generated from standard normal distribution. Effective factors in designing the experiment are the number of explanatory variables $q$, the degree of the correlation among the independent variables $\rho^2$ and the sample size $n$.

Four different values of the correlation $\rho$ corresponding to $0.8, 0.9, 0.99$ and $0.999$ are considered. Moreover, four different values of the number of explanatory variables consisting of $p = 4, 6, 8$ and $12$ are considered in the design of the experiment. The sample size varies as 1200, 500 and 1000. Moreover, we choose the number of principal components using the method of percentage of the total variability which is defined as

$$PTV = \frac{\sum_{j=1}^{r} \lambda_j}{\sum_{j=1}^{p} \lambda_j} \times 100.$$

In the simulation, PTV is chosen as $0.75$ for $p = 4, 8$ and $12$ and $0.83$ for $p = 6$ (see Aguilera et al. (2006)).

The coefficient vector is chosen due to Newhouse and Oman (1971) such that $\beta'\beta = 1$ which is a commonly used restriction, for example see Kibria (2003). We generate the $n$ observations of the dependent variable using the Bernoulli distribution $\text{Be}(\pi_i)$ where $\pi_i = \frac{e^{x_i\beta}}{1+e^{x_i\beta}}$ such that $x_i$ is the $i^{th}$ row of the data matrix $X$.

The simulation is repeated for 2000 times. To compute the simulated MSEs of the estimators, the following equation is used respectively:

$$\text{MSE}(\tilde{\beta}) = \frac{\sum_{r=1}^{2000}(\tilde{\beta}_c - \beta)'(\tilde{\beta}_c - \beta)}{2000} \quad (4.2)$$

where $\tilde{\beta}_c$ is MLE, PCLR, LTL, and PCLTL in the $c^{th}$ replication. The convergence tolerance is taken to be $10^{-6}$.

We choose the biasing parameter as follows:

- LTL: We refer to Asar (2017) and choose $d = \frac{1}{2}min\left\{\frac{\lambda_j}{(1+\lambda_j)}\right\}$ where min is the minimum function and $k_{AM} = \frac{1}{p}\sum_{j=1}^{p}\frac{\lambda_j - d(1+\lambda_j\hat{\alpha}_j^2)}{\lambda_j\hat{\alpha}_j^2}$.

- PCLTL: We use the same estimators used in LTL.

**Table 1. Simulated MSE values of the estimators when $p = 4$**

| n | 200 | | | | 500 | | | | 1000 | | | |
|---|---|---|---|---|---|---|---|---|---|---|---|---|
| $\rho$ | 0.8 | 0.9 | 0.99 | 0.999 | 0.8 | 0.9 | 0.99 | 0.999 | 0.8 | 0.9 | 0.99 | 0.999 |
| MLE | 3.6544 | 7.5668 | 78.5165 | 736.1838 | 3.7399 | 7.1956 | 76.1299 | 782.6464 | 3.7724 | 8.5449 | 71.1559 | 768.2830 |
| LTL | 0.7733 | 0.7048 | 0.5824 | 0.5156 | 0.7725 | 0.7065 | 0.5812 | 0.5128 | 0.7702 | 0.6995 | 0.5719 | 0.5066 |
| PCLR | 1.7587 | 3.0379 | 27.8693 | 226.8247 | 1.7430 | 3.1317 | 25.3154 | 255.0584 | 1.7929 | 3.1873 | 23.3964 | 249.4602 |
| PCLTL | 0.7680 | 0.7003 | 0.5807 | 0.5153 | 0.7660 | 0.6992 | 0.5797 | 0.5125 | 0.7651 | 0.6941 | 0.5698 | 0.5062 |

**Table 2. Simulated MSE values of the estimators when $p = 6$**

| n | 200 | | | | 500 | | | | 1000 | | | |
|---|---|---|---|---|---|---|---|---|---|---|---|---|
| $\rho$ | 0.8 | 0.9 | 0.99 | 0.999 | 0.8 | 0.9 | 0.99 | 0.999 | 0.8 | 0.9 | 0.99 | 0.999 |
| MLE | 6.2446 | 13.4757 | 106.8415 | 1519.4139 | 6.2088 | 13.4665 | 121.5800 | 1230.7433 | 6.5498 | 12.8373 | 129.6968 | 1266.0770 |
| LTL | 0.7787 | 0.7145 | 0.5761 | 0.4817 | 0.7928 | 0.7285 | 0.5649 | 0.4898 | 0.7704 | 0.7202 | 0.5709 | 0.4854 |
| PCLR | 3.1066 | 5.7196 | 46.8706 | 567.2190 | 2.9268 | 5.9565 | 50.5092 | 502.2837 | 3.3763 | 5.5803 | 52.2650 | 478.3712 |
| PCLTL | 0.7771 | 0.7126 | 0.5745 | 0.4814 | 0.7914 | 0.7260 | 0.5636 | 0.4895 | 0.7681 | 0.7184 | 0.5696 | 0.4851 |

**Table 3. Simulated MSE values of the estimators when $p = 8$**

| n     | 200    |         |          |           | 500    |         |          |           | 1000   |         |          |           |
|-------|--------|---------|----------|-----------|--------|---------|----------|-----------|--------|---------|----------|-----------|
| $\rho$ | 0.8    | 0.9     | 0.99     | 0.999     | 0.8    | 0.9     | 0.99     | 0.999     | 0.8    | 0.9     | 0.99     | 0.999     |
| MLE   | 8.3054 | 15.0314 | 184.7005 | 1525.8426 | 8.4735 | 18.6290 | 174.1891 | 1613.8509 | 8.4524 | 17.2693 | 180.3706 | 1641.4217 |
| LTL   | 0.7748 | 0.7407  | 0.5854   | 0.4852    | 0.7811 | 0.7166  | 0.5851   | 0.4756    | 0.7781 | 0.7199  | 0.5920   | 0.4690    |
| PCLR  | 3.9590 | 5.8582  | 59.1569  | 514.4022  | 3.5947 | 7.2542  | 61.6898  | 557.6362  | 3.7627 | 6.8658  | 62.8295  | 565.5316  |
| PCLTL | 0.7689 | 0.7335  | 0.5756   | 0.4820    | 0.7742 | 0.7068  | 0.5729   | 0.4717    | 0.7703 | 0.7088  | 0.5802   | 0.4651    |

**Table 4. Simulated MSE values of the estimators when $p = 12$**

| n     | 200     |         |          |           | 500     |         |          |           | 1000    |         |          |           |
|-------|---------|---------|----------|-----------|---------|---------|----------|-----------|---------|---------|----------|-----------|
| $\rho$ | 0.8     | 0.9     | 0.99     | 0.999     | 0.8     | 0.9     | 0.99     | 0.999     | 0.8     | 0.9     | 0.99     | 0.999     |
| MLE   | 13.7770 | 26.6939 | 263.8436 | 2601.2224 | 11.2652 | 27.3167 | 301.9883 | 3250.0013 | 13.7240 | 25.7586 | 284.4920 | 2623.1785 |
| LTL   | 0.7869  | 0.7270  | 0.6094   | 0.4821    | 0.7925  | 0.7126  | 0.6023   | 0.4740    | 0.7688  | 0.7198  | 0.6255   | 0.4889    |
| PCLR  | 4.6458  | 8.1955  | 75.2659  | 722.7708  | 4.5056  | 8.8706  | 86.6882  | 897.2018  | 5.2046  | 8.5334  | 83.6243  | 809.3412  |
| PCLTL | 0.7796  | 0.7112  | 0.5845   | 0.4716    | 0.7832  | 0.6966  | 0.5766   | 0.4618    | 0.7585  | 0.7038  | 0.5905   | 0.4733    |

According to Tables 1-4, MSE of the MLE is inflated when the degree of correlation is increased. Similarly, if we consider PCLR, its MSE values are also inflated for increasing values of the degree of correlation.

In general, increasing the number of explanatory variables affects the estimators negatively, namely, this situation makes MLE and PCLR less efficient such that MSE of MLE and PCLR increase rapidly. However, LTL and PCLTL are affected slightly when the number of variables changes.

MLE and PCLR produce high MSE values when the sample size is low and the degree of correlation is high. LTLT and PCLTLT are robust to this situation in almost all the cases. Increasing the sample size makes a positive effect on the estimators in most of the situations. However, there is a degeneracy in this property especially when the degree of correlation is high.

LTL and PCLTL are robust to the degree of correlation i.e. increasing the degree of correlation affects the performance of these estimators positively in most of the situations.

Overall, LTL becomes the second-best estimator and the new estimator PCLTL has the lowest MSE value in all the situations considered in the simulation.

## 5. Conclusion

In this paper, we develop a new principal component Liu-type logistic estimator as a combination of the principal component logistic regression estimator and Liu-type logistic estimator to overcome the multicollinearity problem. We have proved some theorems showing the superiority of the new estimator over the other estimators by studying their asymptotic mean squared error matrix criterion. Finally, a Monte Carlo simulation study is presented in order to show the performance of the new estimator. According to the results, it seems that PCLTL is better alternative in multicollinear situations in the binary logistic regression model.

## References

Aguilera, A. M., Escabias, M., and Valderrama, M. J. (2006). Using principal components for estimating logistic regression with high-dimensional multicollinear data. Computational Statistics & Data Analysis, 50(8), 1905-1924.

Akdeniz, F., and Erol, H. (2001). Mean Squared error matrix comparisons of some biased estimator in linear regression. Communications in Statistics-Theory and Methods. 32: 2389-2413

Alheety. M.I., and Kibria, B. M.G. (2013). Modified Liu-Type Estimator Based on (r - k) Class Estimator. Communications in Statistics- Theory and Methods. 2: 304-319.

Asar, Y. (2017). Some new methods to solve multicollinearity in logistic regression. Communications in Statistics-Simulation and Computation, 46(4), 2576-2586.

Asar, Y., and Genç, A. (2016). New Shrinkage Parameters for the Liu-Type Logistic Estimators, Communication in Statistics-Simulation and Computation, 45:3, 1094-1103.

Baye, M. R., and Parkar, F. P. (1984). Comning ridge and pricipal component regression: a money demand illustration. Communications in Statistics-Theory and Methods. 13: 197-225.

Baksalary, J. K., and Trenkler, G. (1991). Nonnegative and positive definiteness of matrices modified by two matrices of rank one. Linear Algebra and its Application. 151:169-184.

Farebrother, R.W. (1976). Further Results on the Mean Square Error of Ridge Regression. Journal of the Royal Statistical Society B, 38,248-250.

Hoerl, A.E., and Kennard, R. W. (1970). Ridge regression: biased estimation for nonorthogonal problems. Technometrics. 12:55-67.

Huang, W.H., Qi, J.J., Huang, N. T. (2009). Liu-type estimator for linear model with linear restrictions. Journal of System Sciences and Mathematical Sciences. 29:937-946

Inan, D., and Erdogan, B. E. (2013). Liu-type logistic estimator. Communications in Statistics-Simulation and Computation, 42(7), 1578-1586.

Kibria, B. M. G. (2003). Performance of some new ridge regression estimators. Communications in Statistics-Simulation and Computation, 32(2), 419-435.

Lee, A. H., and Silvapulle, M. J. (1988). Ridge estimation in logistic regression. Communications in Statistics-Simulation and Computation, 17(4), 1231-1257.

Lesaffre, E., and Max, B.D. (1993). Collinearity in Generalized Linear Regression, Communications in Statistics-Theory and Methods, 22(7), 1933-1952.

Liu, K. (1993). A new class of biased estimate in linear regression. Communications in Statistics- Theory and Methods.22: 393-402.

Li, Y.L., and Yang, H. (2011). Two kinds of restricted modified estimators in linear regression model. Journal of Applied Statistics. 38:1447-1454.

Månsson, K., Kibria, B. G., and Shukur, G. (2012). On Liu estimators for the logit regression model. Economic Modelling, 29(4), 1483-1488.

Newhouse, J. P. and Oman, S. D. (1971). An evaluation of ridge estimators. Rand Corporation(P-716-PR), 1-16.

Rao, C.R., and Toutenburg,H.(1995). Linear models: Least squares and Alternative. Springer-Verlag, New York.

Saleh, A. M. E., and Kibria, B. M. G. (2013). Improved ridge regression estimators for the logistic regression model. Computational Statistics, 28(6), 2519-2558.

Schaefer, R. L., Roi, L. D. and Wolfe, R. A. (1984). A ridge logistic estimator. Communications in Statistics-Theory and Methods, 13(1), 99-113.

Xu, J. W., Yang, H. (2011). On the restricted almost unbiased estimators in linear regression. Journal of Applied Statistics. 38: 605-617.